\documentclass[twocolumn,showpacs,preprintnumbers,amsmath,amssymb,prl]{revtex4}

\usepackage{graphicx}
\usepackage{dcolumn}
\usepackage{bm}
\usepackage{color}
\usepackage{soul}
\usepackage{notes2bib}

\begin{document}



\title{Locking of electron spin coherence over fifty milliseconds  in natural silicon carbide}

\author{D.~Simin$^{1}$}
\author{H.~Kraus$^{2,1}$}
\author{A.~Sperlich$^{1}$}
\author{T.~Ohshima$^{2}$}
\author{G.~V.~Astakhov$^{1}$}
\email[E-mail:~]{astakhov@physik.uni-wuerzburg.de}
\author{V.~Dyakonov$^{1,3}$}
\email[E-mail:~]{dyakonov@physik.uni-wuerzburg.de}

\affiliation{$^1$Experimental Physics VI, Julius-Maximilian University of W\"{u}rzburg, 97074 W\"{u}rzburg, Germany \\
$^2$National Institutes for Quantum and Radiological Science and Technology, Takasaki, Gunma 370-1292, Japan \\
$^3$Bavarian Center for Applied Energy Research (ZAE Bayern), 97074 W\"{u}rzburg, Germany}

\begin{abstract}
We demonstrate that silicon carbide (SiC) with natural isotope abundance can preserve a coherent spin superposition in silicon vacancies over unexpectedly long time approaching 0.1~seconds. The spin-locked subspace with drastically reduced decoherence rate is attained through the suppression of heteronuclear spin cross-talking by applying a moderate magnetic field in combination with dynamic decoupling from the nuclear spin baths.  We identify several phonon-assisted mechanisms of spin-lattice relaxation, ultimately limiting quantum coherence, and find that it can be extremely long at cryogenic temperature, equal or even longer than 8~seconds. Our approach may be extended to other polyatomic compounds  and open a path towards improved qubit memory for wafer-scale quantum techmologies. 
\end{abstract}

 
\date{\today}

\maketitle

\textit{Introduction} --- Long electron quantum coherence in solid-state systems is the ultimate prerequisite for new technologies based on quantum phenomena \cite{Ladd:2010kq, Awschalom:2013in}. Particularly, the sensitivity of quantum sensors scales with the electron spin coherence time  $T_2$  \cite{Taylor:2008cp, Balasubramanian:2009fu}.  One of the common sources of decoherence is the interaction with fluctuating nuclear spins, and the usual way to prolong spin coherence is to perform isotope purification of the crystal. Indeed, the longest electron $T_2$ times of about $1 \, \mathrm{s}$  and $0.6 \, \mathrm{s}$ have been reported for spin-free silicon $^{28}$Si and diamond $^{12}$C crystals, respectively \cite{Tyryshkin:2011fi, BarGill:2013dq}. However, isotope purification is a technologically demanding procedure, which is not always possible. Therefore, one of the key challenges in quantum information science is to achieve long-lived spin coherence in natural materials. 

To address this goal, we combine two approaches. First, we exploit the suppression of mutual spin flip-flop processes between different types of nuclei in binary compounds, which occur  in strong enough magnetic fields according to the theoretical simulations of Ref.~\cite{Yang:2014kqa}. Second, we use a periodic train of radiofrequency (RF) pulses to refocus spin coherence and decouple electron spins from inhomogeneous environment, similar to that applied for color centers \cite{BarGill:2013dq} and quantum dots (QDs) \cite{Greilich:2006bf}. 

In recent years, SiC is attracting continuously growing interest as a technologically perspective platform for quantum spintronics \cite{Baranov:2011ib, Koehl:2011fv, Riedel:2012jq, Soltamov:2012ey, Kraus:2013di, Kraus:2013vf, Carter:2015vc, Falk:2015iz, Klimov:2015bm} with the ability for single spin engineering and control \cite{Castelletto:2013el, Christle:2014ti, Widmann:2014ve, Fuchs:2015ii}. The longest  $T_2$ in SiC reported to date is $1 \, \mathrm{ms}$ at cryogenic temperature \cite{Christle:2014ti}. We observe that in a finite magnetic field, a coherent spin superposition  can be locked over longer time, which continuously increases  up to about $75 \, \mathrm{ms}$ with the number of decoupling pulses. The absence of saturation indicates that the longest possible spin locking time $T_2^{\mathrm{SL}}$ is not reached in our experiments. We estimate the saturation level to be approximately $T_2^{\mathrm{SL}} \rightarrow 0.3 \, \mathrm{s}$, which can be viewed as the upper boundary for $T_2$ \cite{BarGill:2013dq, Naydenov:2011eo}. Although we  demonstrate the preservation of one spin component only, this spin locking can still be used for quantum sensing with extraordinarily sensitivity \cite{Naydenov:2011eo, deLange:2011joa, Hirose:2012id, Loretz:2013io}. Remarkably, the observed $T_2^{\mathrm{SL}}$ time is within the same order of magnitude with the  $T_2$ record values in the solid state but achieved without isotope purification of the crystal.  

\begin{figure*}[t]
\includegraphics[width=.69\textwidth]{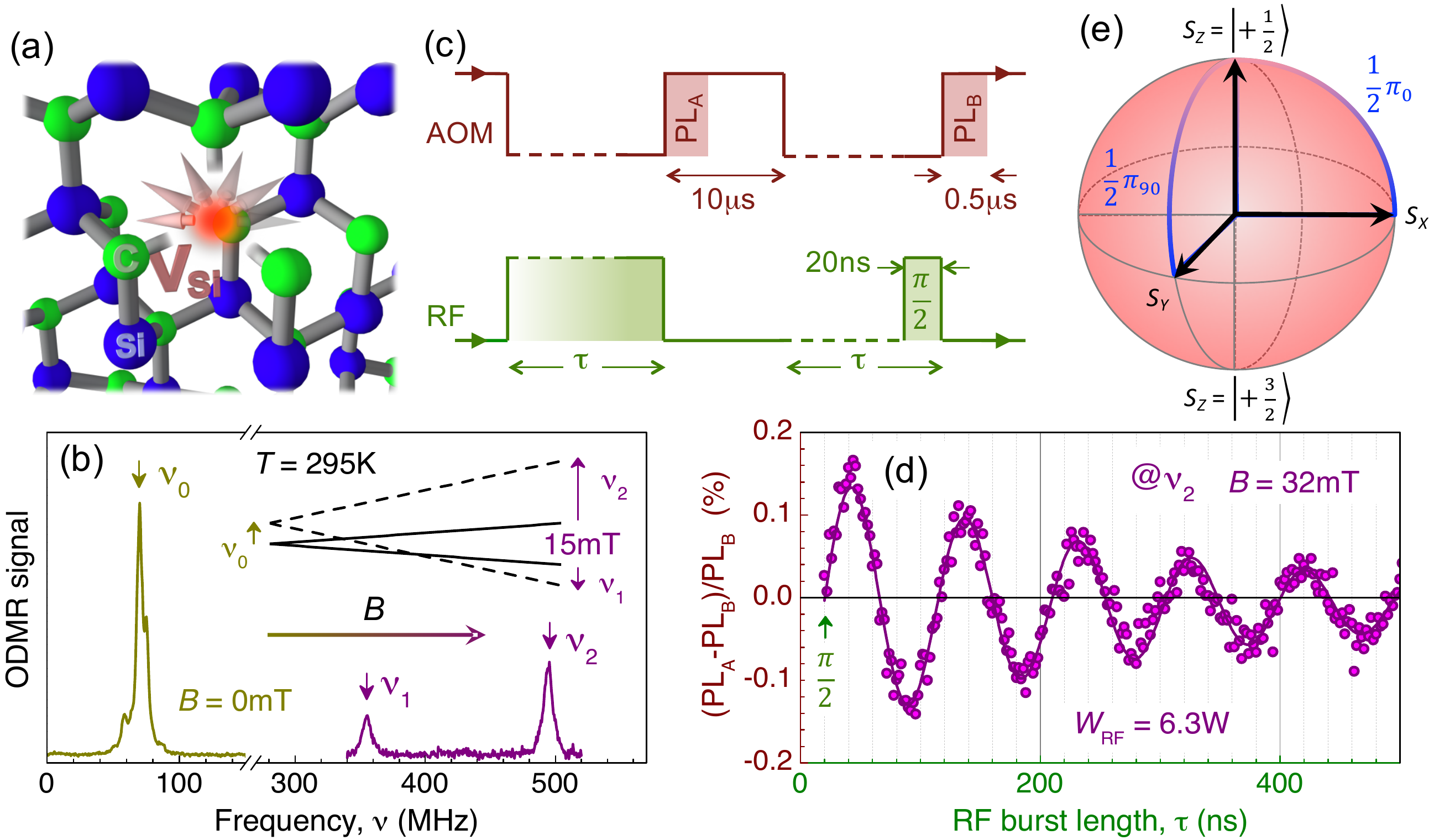}
\caption{Room temperature coherent control of silicon vacancies in 4H-SiC. (a) A schematic representation of the $\mathrm{V_{Si}}$ spin under optical pumping. (b) ODMR spectra in zero magnetic field and in a magnetic field of $15 \, \mathrm{mT}$. The inset shows the $\mathrm{V_{Si}}$  spin-3/2 splittings with growing magnetic field. The solid (dashed) lines represent the pumped $m_S = \pm 1/2$ (depleted $m_S = \pm 3/2$) states. The vertical arrows indicate the RF induced spin transitions. (c) The AOM (pump laser) and RF switch  sequences to observe Rabi oscillations and to calibrate the $\pi / 2$ pulse.  (d) Exemplary Rabi oscillations between the $m_S = + 1/2$ and $m_S = + 3/2$ states at $B = 32 \, \mathrm{mT}$.  The solid line is a fit to Eq.~(\ref{Rabi}). (e) Definition of different spin states on the Bloch sphere. } \label{fig1}
\end{figure*}

\textit{Coherent control} ---   All results reported here are obtained using a commercial 4H-SiC wafer. We use electron irradiation to generate silicon vacancies ($\mathrm{V_{Si}}$) [Fig.~\ref{fig1}(a)] \cite{Supplemental_Material}. They possess spin $S = 3 /2$ ground state \cite{Kraus:2013di}, which  is split in two sublevels $m_S = \pm 1/2$ and $m_S = \pm 3/2$ in zero magnetic field. We concentrate on the V2-type $\mathrm{V_{Si}}$ with the zero-field splitting $2 D  = 70 \, \mathrm{MHz}$. Due to optical pumping and spin-dependent photoluminescence (PL), $2D$ is directly seen in optically detected magnetic resonance (ODMR) experiments  of Fig.~\ref{fig1}(b) \cite{Kraus:2013vf}. In an external magnetic field $B$, the spin degeneracy is lifted up resulting in four distinct resonances \cite{Simin:2016cp,Simin:2015dn}. Here, we concentrate on the $\nu_1$ and $\nu_2$ ODMR lines [Fig.~\ref{fig1}(b)], corresponding to the transitions $(-1/2 \rightarrow -3/2)$ and $(+1/2 \rightarrow +3/2)$, respectively. 

Coherent spin manipulation is performed by means of the pulsed ODMR technique \cite{Supplemental_Material}. In order to observe Rabi oscillations, we vary the duration $\tau$ of the first RF pulse and fix the duration of the second (reference) pulse to $20 \, \mathrm{ns}$ [Fig.~\ref{fig1}(c)]. We drive the $\nu_2$ transition and the spin polarization, measured as $\mathcal{S} = \mathrm{(PL_A - PL_B ) / PL_B}$, oscillates with $\tau$ as [Fig.~\ref{fig1}(d)] 
\begin{equation}
\mathcal{S}(\tau) = - \frac{1}{2}  \frac{ \Delta_{\mathrm{PL}}}{\mathrm{PL}}  \, e^{- \tau / T_R} \cos  \Omega_R \tau  + C_{\pi/2}   \,.
\label{Rabi}
\end{equation}
Here, the decay time is $T_R = 309 \pm 13 \, \mathrm{ns}$ and $\Omega_R$ is the angular Rabi frequency, which depends on the driving RF power $W_{\mathrm{RF}}$. The ODMR contrast is $\Delta_{\mathrm{PL}} / \mathrm{PL} = 0.32 \%$. 
For $W_{\mathrm{RF}} = 6.3 \, \mathrm{W}$ we obtain $C_{\pi/2} = 0$, meaning that the reference pulse in Fig.~\ref{fig1}(c) is equal to $\pi / 2$. 

\begin{figure}[!]
\includegraphics[width=.48\textwidth]{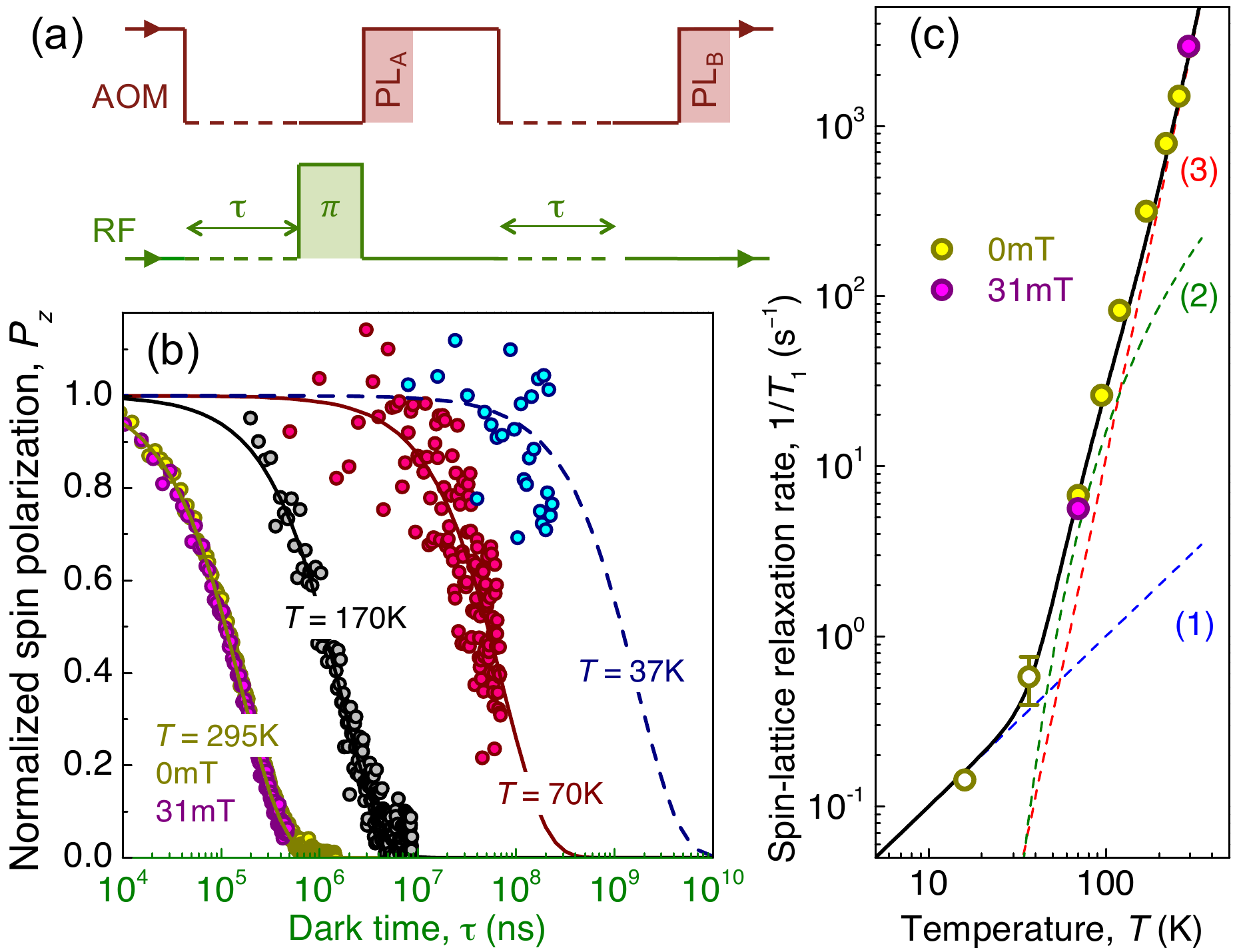}
\caption{Temperature dependence of the spin-lattice relaxation time.  (a) A pulse sequence to measure $T_1$ of the $\mathrm{V_{Si}}$ spins. (b)  Measurements of $T_1$ at different temperatures in $B = 0 \, \mathrm{mT}$ and  $B =31 \, \mathrm{mT}$. The solid lines are fits to $P_z = \exp (-2\tau/T_1)$  \cite{Widmann:2014ve}.  The dashed line is an exponential fit using the linear part of the decay for $\tau \ll T_1 $. (c) Spin-lattice relaxation rate $1 / T_1$ as a function of temperature.  The solid and dashed lines represent a fit to Eq.~(\ref{T1-Temp}). The values of fitting parameters are summarized in table ~\ref{Aparam-Temp}. 
} \label{fig2}
\end{figure}

\textit{Spin-lattice relaxation} --- Having calibrated the pulse duration, we first measure the spin-lattice relaxation time $T_1$, which determines the absolute limit for spin coherence. Figure~\ref{fig2}(a) shows the pulse sequence to obtain $T_1$, and exemplary measurements for different temperatures  are presented in Fig.~\ref{fig2}(b). Each experimental curve is fitted to a single-exponential function, yielding $T_1$. We find no variation of $T_1$ with $B$ as reported for the NV centers in diamond  \cite{Jarmola:2012co}, at least up to $31  \, \mathrm{mT}$.  The fit of Fig.~\ref{fig2}(b) at room temperature yields $T_1 = 340 \pm 30  \, \mathrm{\mu s}$, in agreement with earlier studies  \cite{Riedel:2012jq}. With decreasing temperature, $T_1$ increases by several orders of magnitude. At the lowest temperature $T = 17 \, \mathrm{K}$, the spin polarization preserves over  90\% of its initial value in a timespan of $1 \, \mathrm{s}$, and using the linear part of the exponential decay we estimate $T_1 \approx  8 \, \mathrm{s}$. 

\begin{figure*}[!]
\includegraphics[width=.69\textwidth]{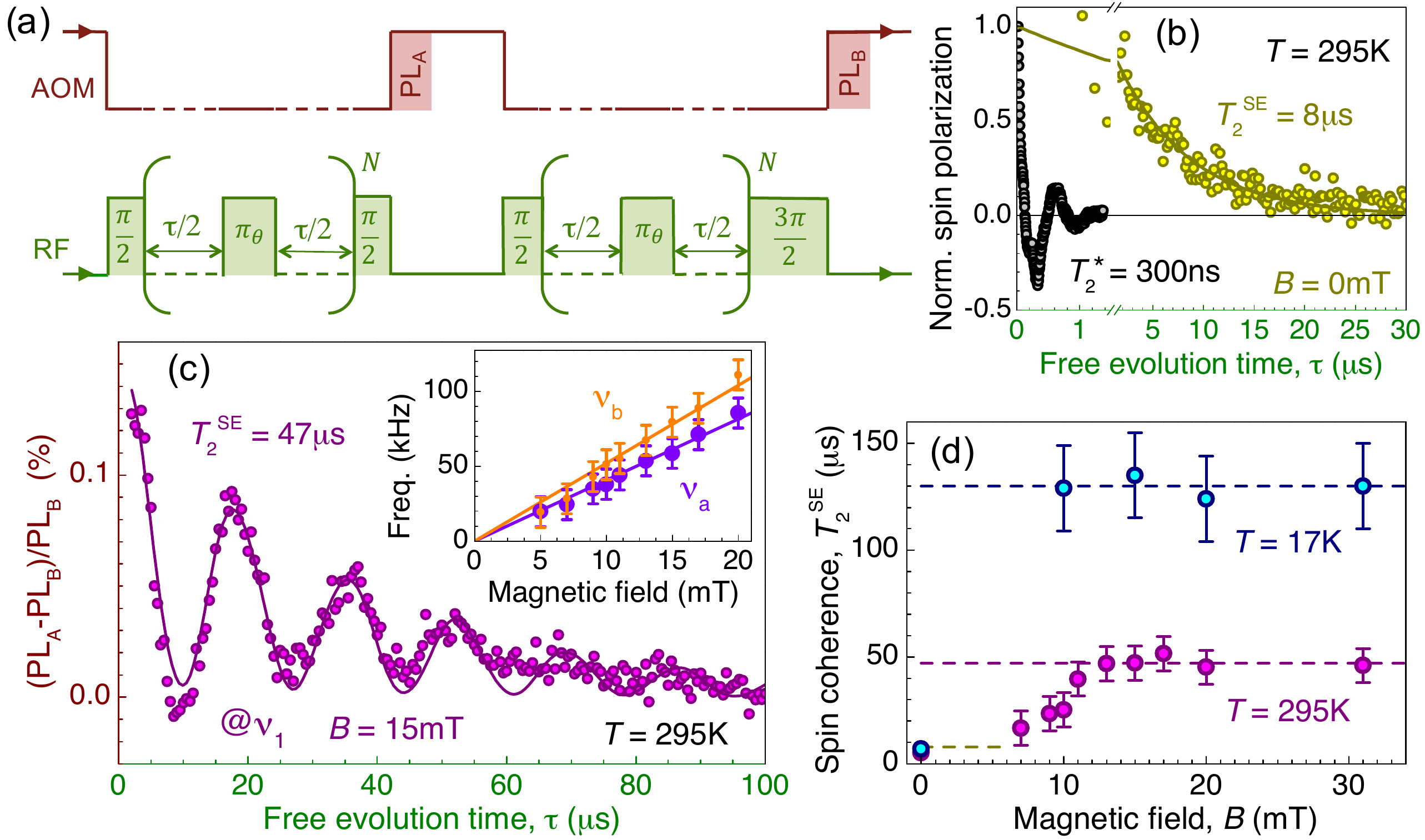}
\caption{Spin-echo coherence time. (a) The pulse sequences to perform Ramsey ($N = 0$), SE ($N=1$ and $\theta = 0^{\circ}$) and CPMG ($N>1$ and $\theta = 90^{\circ}$) measurements. Here, $N$ is the number of $\pi_\theta$ pulses with $\theta$ denoting the phase shift relative to the first $\pi/2$ pulse. (b) Ramsey and SE decay curves in zero magnetic field at room temperature.  The solid line is a fit to  $\exp (-\tau/T_2^{\mathrm{SE}})$. (c) Room temperature SE decay curve in a magnetic field of $15 \, \mathrm{mT} $, recorded at the $\nu_{1}$ resonance. The solid line is a fit to Eq.~(\ref{ESEEM}). The inset represents the Larmor frequencies as a function of the magnetic field. Linear fits of these data give slopes $\nu_a = 4.1 \pm 0.9 \, \mathrm{kHz / mT}$ and $\nu_b = 5.2 \pm 0.9 \, \mathrm{kHz / mT}$. The size of the symbols represents the modulation depth. (d) SE coherence time as a function of the magnetic field at different temperatures. The horizontal dashed lines are to show the saturation levels.} \label{fig3}
\end{figure*}

To gain insight into the spin-lattice relaxation processes in SiC, we plot in Fig.~\ref{fig2}(c) the relaxation rate $1/T_1$ against $T$ in  log-log scale. Our experimental data can be perfectly fitted to a power law function 
\begin{equation}
\frac{1}{T_1 (T)}= A_0 +A_1 T + A_5 T^5 + \frac{R}{e^{\Delta / k_B T} - 1}    \,,
\label{T1-Temp}
\end{equation}
which describes various phonon-assisted spin relaxation mechanisms, similar to that reported for silicon and diamond \cite{Tyryshkin:2011fi,Jarmola:2012co}. For $T > 120 \, \mathrm{K}$, thermally excited high-energy phonons result in spin-lattice relaxation via a two-phonon Raman process  $1/T_1 \approx A_5 T^5$ [dashed line (3) in Fig.~\ref{fig2}(c)] \cite{Abragam_Bleaney}. At intermediate temperatures, an Orbach-like process through a vibrational state may become significant  $1/T_1 \propto (e^{\Delta / k_B T} - 1)^{-1}$ [dashed line  (2) in Fig.~\ref{fig2}(c)] \cite{Abragam_Bleaney}. Here, $\Delta \approx 25 \, \mathrm{meV}$ corresponds to the energy of the local phonon mode at the $\mathrm{V_{Si}}$ defect, which can be roughly estimated from the separation between the zero-phonon line and phonon sideband maximum in the PL spectrum  \cite{Fuchs:2015ii}. For  $T < 30 \, \mathrm{K}$, the relaxation through single phonon scattering should be taken into account  $1/T_1 \approx A_1 T$ [dashed line (1) in Fig.~\ref{fig2}(c)] \cite{Abragam_Bleaney}. At cryogenic temperatures, the cross relaxation with residual impurity/defect spins may also play a role. This spin relaxation mechanism is temperature independent, and from the experimental data of Fig.~\ref{fig2}(c) we estimate the longest  achievable spin-lattice relaxation time to be at least $T_1 (0) = 1 / A_0 \geqslant 8 \, \mathrm{s}$. 

\begin{table}[tdtp]
\caption{The parameters of spin-phonon interaction found from the fit of the experimental data of Fig.~\ref{fig2}(c) to Eq.~(\ref{T1-Temp}). We assume for simplicity $A_0 \leqslant 0.125 \, \mathrm{s^{-1} }$ and $\Delta = 25 \, \mathrm{meV}$. }
\begin{center}
\begin{tabular}{|c|c|c|}
$A_1$ $(\mathrm{K^{-1} s^{-1} })$ & $A_5$ $(\mathrm{K^{-5} s^{-1} })$ & $R$ $(\mathrm{s^{-1} })$ \\
$1.0 \pm 0.2 \times 10^{-2}$ & $1.1 \pm 0.2 \times 10^{-9}$  & $300 \pm 150$  \\
\end{tabular}
\end{center}
\label{default}
\label{Aparam-Temp}
\end{table}

\textit{Spin-echo coherence} --- 
The generalized pulse sequence to measure the spin coherence time is presented in Fig.~\ref{fig3}(a). It is instructive to start with Ramsey interferometry of Fig.~\ref{fig3}(b), giving the inhomogeneous spin coherence time 
$T_2^* \approx 300  \, \mathrm{ns}$. 
In spin echo experiments with an additional refocusing ($\pi$) pulse, the spin coherence time can be significantly prolonged. In $B = 0 \, \mathrm{mT}$, we obtain the spin-echo coherence time  $T_2^{\mathrm{SE}} = 8 \pm 2  \, \mathrm{\mu s}$ [Fig.~\ref{fig3}(b)]. Upon application of an external magnetic field $B \| c$ [Fig.~\ref{fig3}(c)], we observe electron spin echo envelope modulation (ESEEM) \cite{ Koehl:2011fv,Christle:2014ti,Widmann:2014ve,Carter:2015vc}. In case of the $\nu_1$ transition,  it is well fitted to 
\begin{equation}
\mathcal{S}(\tau) = \frac{ \Delta_{\mathrm{PL}}}{\mathrm{PL}}  \, e^{- \tau / T_2^{\mathrm{SE}}}  \prod_{j=a,b}  \left( 1 - K_j \sin^2  \left( \pi \nu_j \tau  \right)  \right)   \,
\label{ESEEM}
\end{equation}
with two frequencies $\nu_a$ and $\nu_b$, shifting linear with $B$ [the inset of Fig.~\ref{fig3}(c)]. 
Similar frequencies were observed for the divacancy ESEEM in SiC and associated with the Larmor precession of  $^{29}$Si and $^{13}$C nuclear spins \cite{ Koehl:2011fv}. Remarkably, the ESEEM for the $\nu_2$ transition shows more complex behavior \cite{Supplemental_Material} and agrees with the earlier reported measurements on silicon vacancies in SiC \cite{Carter:2015vc}. The origin of this difference is unclear.  Given that the applied magnetic field $B = 15 \, \mathrm{mT}$ lies in the vicinity of the excited-state level anticrossing \cite{Carter:2015vc,Simin:2016cp},  the observed asymmetry in the ESEEM may indicate dynamic nuclear spin polarization \cite{Falk:2015iz}. 

From the fits to Eq.~(\ref{ESEEM}) \cite{Supplemental_Material}, we also obtain $T_2^{\mathrm{SE}}$ as a function of $B$ [Fig.~\ref{fig3}(d)]. At room temperature, the spin coherence time increases with growing magnetic field and saturates for $B > 11 \, \mathrm{mT}$ at a level $T_2^{\mathrm{SE}} = 47 \pm 8  \, \mathrm{\mu s}$. This value is slightly shorter than the spin coherence time of single $\mathrm{V_{Si}}$  centers \cite{Widmann:2014ve}, and such a magnetic field dependence is in a qualitative agreement with the theoretical prediction \cite{Yang:2014kqa}, though the $T_2$ saturation is expected for higher fields. The measurements of Figs.~\ref{fig3}(a)-\ref{fig3}(c) are repeated at a temperature $T = 17  \, \mathrm{K}$ \cite{Supplemental_Material} and depict qualitatively the same field dependence  (Fig.~\ref{fig3}d). Without external magnetic field, $T_2^{\mathrm{SE}}$ is temperature independent, and for $B > 11 \, \mathrm{mT}$ the spin coherence time saturates at a higher level $T_2^{\mathrm{SE}} = 130 \pm 20  \, \mathrm{\mu s}$.

\begin{figure}[t]
\includegraphics[width=.48\textwidth]{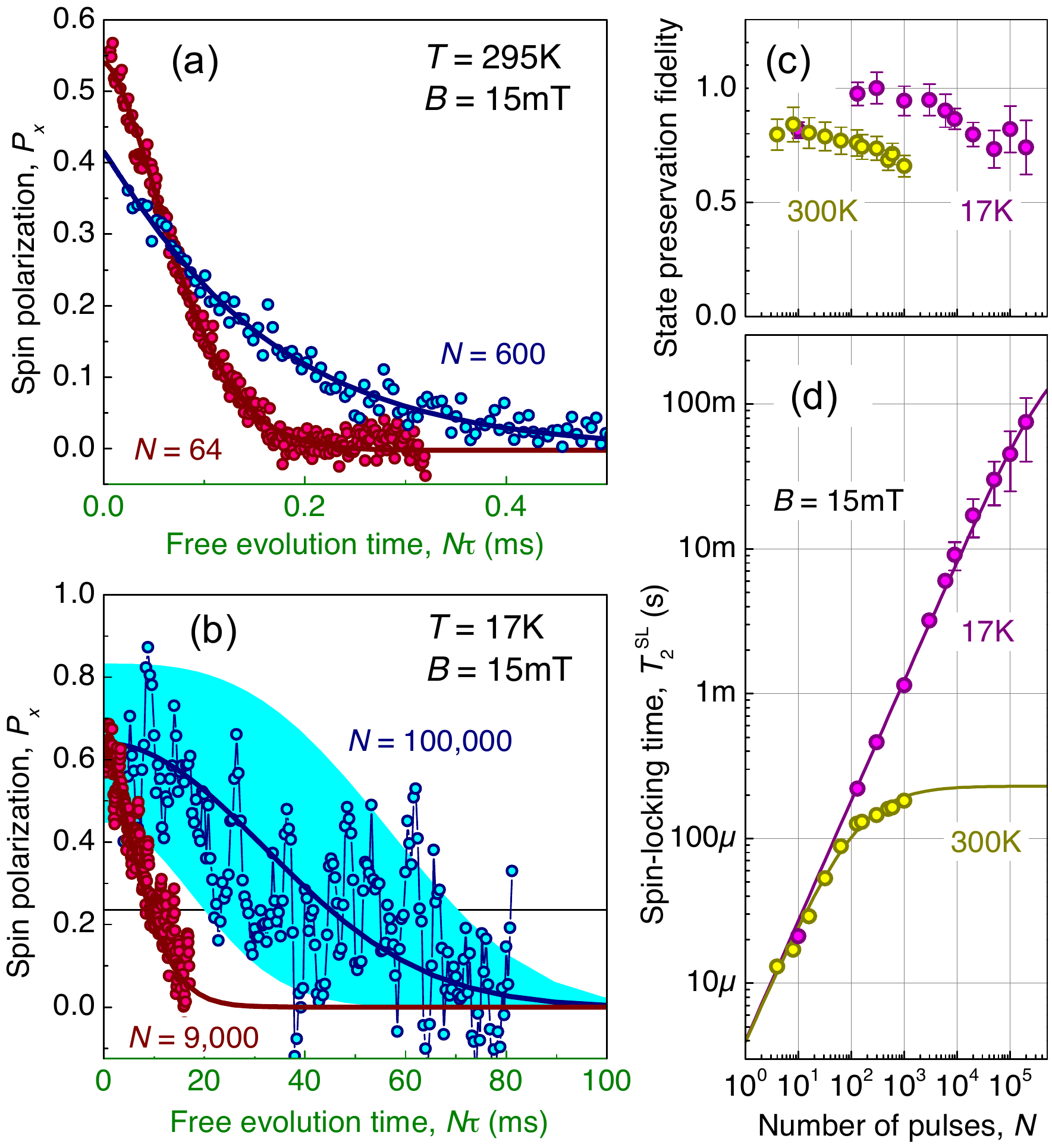}
\caption{Spin locking using dynamic decoupling.  (a)  Room temperature measurements of the spin-locking time $T_2^{\mathrm{SL}}$ using CPMG pulse sequences, performed for different numbers ($N$) of $\pi_{90}$ pulses. The solid lines are fits to stretched exponent in the form $P_x = (2 \mathcal{F_S} - 1) \exp \left[ -(N \tau/T_2^{\mathrm{SL}})^n \right]$, with $ \mathcal{F_S}$, $T_2^{\mathrm{SL}}$ and $n$ being free parameters.  (b)  The same as (a), but measured at $T = 17 \, \mathrm{K}$.   The light area represents a range of the best fits within the standard deviation for $\mathcal{F_S}$ and $T_2^{\mathrm{SL}}$. The horizontal line indicates the $(2 \mathcal{F_S} - 1) e^{-1}$ level. (c) State preservation fidelity $\mathcal{F_S}$ as a function of $N$ at  two different temperatures, obtained from the quantum state tomography in the limit $\tau \rightarrow 0$.  (d)  Spin-locking time $T_2^{\mathrm{SL}}$  as a function of $N$ for two different temperatures ($T = 295 \, \mathrm{K}$ and $T = 17 \, \mathrm{K}$). The solid lines are fits to the power function of Eq.~(\ref{T2-N}). At $T = 17 \, \mathrm{K}$,   the spin-locking time follows $T_2^{\mathrm{SL}} (N) = T_2^{\mathrm{SL}} (0) N^\kappa$ with $T_2^{\mathrm{SL}} (0) = 4 \pm 1 \, \mathrm{\mu s}$ and $\kappa = 0.83 \pm 0.05$.} \label{fig4}
\end{figure}

\textit{Locking of spin coherence} --- In order to preserve a coherent state even longer as in the spin-echo experiments, we dynamically decouple the $\mathrm{V_{Si}}$ spins from the $^{29}$Si and $^{13}$C nuclear spin baths at $15 \, \mathrm{mT}$ [Figs.~\ref{fig4}(a) and \ref{fig4}(b)]. We choose the Carr-Purcell-Meiboom-Gill (CPMG) pulse sequence for this purpose  \cite{Meiboom:1958dq}, which is successfully applied for the NV defects in diamond \cite{deLange:2010ga,BarGill:2013dq}. Multiple ($N>1$) $\pi_\theta$ pulses repetitively refocus spin coherence and the phase shift $\theta = 90^{\circ}$ makes the CPMG protocol robust against pulse uncertainties [Fig.~\ref{fig3}(a)]. With increasing number of $\pi_{90}$ pulses, the ESEEM modulation depth decreases and simultaneously the non-oscillating contribution in $\mathcal{S}$ rises up. For $N>15$, the ESEEM pattern is not observed any more, indicating that the $^{29}$Si and $^{13}$C nuclear spin baths are now  decoupled. 

In contrast to spin echo, the CPMG pulse sequence provides single-axis dynamic decoupling. This means that only one spin component $S_x = (|+1/2 \rangle  + |+3/2 \rangle )/\sqrt{2}$ is preserved, while the perpendicular component $S_y = (|+1/2 \rangle  + i |+3/2 \rangle )/\sqrt{2}$ decays fast. In other words, the $\mathrm{V_{Si}}$ spin is locked along the $x$ direction on the Bloch sphere [Fig.~\ref{fig1}(e)]. In oder to measure the corresponding spin-locking time $T_2^{\mathrm{SL}}$ and distinguish it from the spin-echo time $T_2^{\mathrm{SE}}$, we first justify that $S_x$ is indeed preserved even after a large number of decoupling pulses. We perform quantum state tomography  \cite{Supplemental_Material}, demonstrating that the state preservation fidelity $\mathcal{F_S}$  varies between $1$ and $0.7$ for all $N$ in the limit $\tau \rightarrow 0$ [Fig.~\ref{fig4}(c)]. With increasing free evolution time $N \tau$, the spin polarization  decreases and can be well fitted to $P_x = (2 \mathcal{F_S} - 1) \exp \left[ -(N \tau/T_2^{\mathrm{SL}})^n \right]$  \cite{Supplemental_Material}, as shown in Figs.~\ref{fig4}(a) and \ref{fig4}(b). 

We first analyze $T_2^{\mathrm{SL}}$ against $N$ at room temperature $T = 295 \, \mathrm{K}$  shown in Fig.~\ref{fig4}(d). The spin-locking time $T_2^{\mathrm{SL}} (N)$ increases monotonically up to $N = 100$, following a power law in the form 
\begin{equation}
\frac{1}{T_2^{\mathrm{SL}} (N)} = \frac{1}{T_2^{\mathrm{SL}} (0) N^\kappa} + \frac{1}{ T_2^{\mathrm{SL}} (\infty)} \,.
\label{T2-N}
\end{equation}
Here, the scaling factor $\kappa = 0.83 \pm 0.05$ slightly deviates from the expected general scaling with $\kappa = 2/3$ \cite{deLange:2010ga} and depends on the spin-bath dynamics \cite{BarGill:2012eq}. The best fit suggests the longest spin locking time for $N \gg$ 100 $T_2^{\mathrm{SL}} (\infty) = 230  \pm 30 \, \mathrm{\mu s}$. It is ultimately limited by the spin-lattice relaxation time $T_1$. 

We now discuss experiments at $T = 17 \, \mathrm{K}$ presented in Fig.~\ref{fig4}(b). The spin-locking time $T_2^{\mathrm{SL}} (N) $ continuously increases with $N$ [Fig.~\ref{fig4}(d)].  For $200,000$ pulses, it reaches $T_2^{\mathrm{SL}} (200,000) = 75 \pm 35 \, \mathrm{ms}$ with the state preservation fidelity $\mathcal{F_S} = 0.74 \pm 0.12$ (Fig.~\ref{fig4}c). The relatively large errors are caused by a consequently required longer integration time with increasing $T_2^{\mathrm{SL}}$, which is our limitation at present.  Using resonant optical excitation \cite{Riedel:2012jq}, one can achieve significantly higher ODMR contrast \cite{Baranov:2011ib}, and spin locking should be feasible for $N > 200,000$ refocusing pulses. The experimental points at $T = 17 \, \mathrm{K}$ of Fig.~\ref{fig4}(d) are also well fitted to Eq.~(\ref{T2-N}) with the same $T_2^{\mathrm{SL}} (0)$ and $\kappa$ as at room temperature. Using Eq.~(\ref{T2-N}), we extrapolate $T_2^{\mathrm{SL}} (\infty) \rightarrow 0.3  \, \mathrm{s}$ for $N > 10^7$. 

To completely characterize the decoherence process in the $\mathrm{V_{Si}}$ defect subject to the CPMG decoupling protocol, we perform quantum process tomography  \cite{Chuang:1997jx, Supplemental_Material}. The process matrices are found to be diagonal within our experimental error ($\pm 0.05$) and application of the Pauli twirling approximation  \cite{Ghosh:2012ec, Geller:2013jr, Supplemental_Material} points at a spin-locked subspace with an exceptionally long decoherence time in accord with theoretical considerations \cite{Zhang:2007hf, Ridge:2014fs, Wang:2012hl}.  Our experiments on SiC hence demonstrate a 75-fold improvement of the spin coherence associated with one particular component compared to the earlier reported results on this material \cite{Christle:2014ti}.  

\textit{Discussion} --- For actual quantum information processing with arbitrary initial and final spin states, one could use two-axis sequences like the XY-family, which preserve all spin components equally \cite{deLange:2010ga}. However, these sequences are very sensitive to flip angle errors  due to imperfect pulse lengths \cite{Shim:2012ha}. We observe that after hundred XY-4 pulses our measurements become unusable. In case of the CPMG decoupling protocol with single rotation axis, the spin component along this axis preserves its value for thousands of pulses and describes a near-ideal limit for the $T_2$ coherence time \cite{BarGill:2013dq, Wang:2012ej}. Using composite pulse sequences \cite{Souza:2011kd}, like in the Knill dynamic decoupling (KDD) protocol, arbitrary quantum states can be preserved (as for the XY-family) while keeping robustness against pulse errors (as for the CPMG pulse sequence).  


Our measurements are restricted to SiC, but we expect similar behavior in other binary (or polyatomic) compounds where spin-carrying nuclear isotopes are not highly abundant, particularly, transition/alkali metal impurities in ZnO \cite{Tribollet:2008jn}, MgO \cite{Bertaina:2009fa} and ZnSe \cite{Greilich:2012je}. 
 
\begin{acknowledgments}
This work has been supported by the German Research Foundation (DFG) under grant AS 310/4, by the ERA.Net RUS Plus program and the German Federal Ministry of Education and Research (BMBF) within project DIABASE as well as JSPS Grant-in-Aid for Scientific  Research (B) 26286047. HK acknowledges the support of the German Academic Exchange Service (DAAD), with funds of the BMBF and EU Marie Curie Actions (DAAD P.R.I.M.E. project 57183951). 
\end{acknowledgments}




\end{document}